\documentclass[aps,prl,superscriptaddress, twocolumn,showpacs]{revtex4}
\usepackage{dcolumn}
\usepackage{bm}
\usepackage[dvips]{graphics}
\begin{document}
Phys. Rev. Lett {\bf 96}£¬ 140401 (2006)
\title{
Bose-Einstein Condensation with Entangled Order Parameter}
\author{Yu Shi}
\thanks{Corresponding author.  \\Electronic address:
yushi@fudan.edu.cn} \affiliation{Department Physics, Fudan
University, Shanghai 200433, China}
\author{Qian Niu}
\affiliation{Department of Physics, University of
Texas, Austin, TX 78712}

\begin{abstract}
We propose a practically accessible non-mean-field ground state of
Bose-Einstein condensation (BEC), which occurs in an interspecies
two-particle entangled state, and is thus described by an entangled
order parameter. A suitably defined entanglement entropy is used as
the characterization of the non-mean-field nature, and is found to
persist in a wide parameter regime.  The interspecies entanglement
leads to novel interference terms in the dynamical equations
governing the single particle orbital wavefunctions. Experimental
feasibility and several methods of probe are discussed. We urge the
study of multi-channel scattering between different species of
atoms.
\end{abstract}


\maketitle

For a mixture of two distinct quantum many-particle systems, the
usual approach is to associate an order parameter to each. For
example, in a two-component BEC~\cite{ho0,myatt,bloch,modugno}, each
component is separately described by a condensate wavefunction,
although the two wavefunctions are classically coupled. Beyond this
approach is BEC of spin-$1$~\cite{ho,zhou,stenger,leggett} or
spin-$\frac{1}{2}$ atoms~\cite{kuklov,lobo2}, but under
consideration is one many-particle system in which all the atoms are
indistinguishable and each atom can jump in between different spin
states. In this Letter, we propose a peculiar mixture of two {\em
distinct} species of atoms $a$ and $b$, which represents a novel
situation different from both cases above. Each atom can flip
between two pseudo spin states $\uparrow$ and $\downarrow$, but
cannot transit between $a$ and $b$. The total occupation number of
each species is conserved, but that of each spin state of each
species is not conserved because of scattering between different
species.  It is shown that the ground state is BEC of  inter-species
pairs in a two-particle state which, as the global BEC order
parameter, is quantum-entangled. As such, this work extends the
notion of entanglement to  order parameters of many-body systems,
and represents a direction different from other works on
entanglement in BEC~\cite{lit}. In particular, the entanglement
between the two species can serve as a characterization of the
non-mean-field nature. The detection methods are discussed. We also
derive the dynamical equations of the single particle orbital
wavefunctions, each  containing an interference term which is absent
in BEC mixtures studied previously. As a condensate of nonlocal Bose
pairs, our system is also related to the condensation of bound
molecules and of paired fermions~\cite{mol}. Experimentally, the two
pseudo spin states can be implemented as two hyperfine $m_F$ states
in a magnetic trap. The atoms can be constrained in the two pseudo
spin states by exploiting the conservations of energy and of the
total $m_F$ in a scattering. Because of the implemental feasibility
as well as the persistence of the entanglement in a wide parameter
regime, this BEC system represents a {\em practically accessible}
non-mean-field ground state, in contrast with previous BEC models.

In ignorance of the depletion, the orbit of each atom, of species
$i$ ($i=a,b$) and with pseudo spin $\sigma$, is constrained in the
single-particle ground state $\phi_{i\sigma}(\mathbf{r}_i)$. The
Hamiltonian can be written as ${\cal H}  = \sum_{\sigma} f_{i\sigma}
N_{i\sigma}+
\frac{1}{2}\sum_{\sigma\sigma'}K^{(ii)}_{\sigma\sigma'}N_{i\sigma}N_{i\sigma'}
 +\sum_{\sigma\sigma'}K^{(ab)}_{\sigma\sigma'}N_{a\sigma}
N_{b\sigma'} + \frac{K_e}{2} (a^{\dagger}_{\uparrow}a_{\downarrow}
b^{\dagger}_{\downarrow}b_{\uparrow} +
a^{\dagger}_{\downarrow}a_{\uparrow}
b^{\dagger}_{\uparrow}b_{\downarrow}),$ where  $a_{\sigma}$ and
$b_{\sigma}$ are annihilation operators,
$N_{a\sigma}=a^{\dagger}_{\sigma}a_{\sigma}$,
$N_{b\sigma}=b^{\dagger}_{\sigma}b_{\sigma}$,
$K^{ij}_{\sigma\sigma'}$'s are  related to scattering
lengths~\cite{scatter}, $f_{i\sigma} \equiv \epsilon_{i\sigma}-
K^{(ii)}_{\sigma\sigma}/2$, where $\epsilon_{i\sigma}$ is the single
particle energy.
$\epsilon_{a\uparrow}-\epsilon_{a\downarrow}=\epsilon_{b\downarrow}-\epsilon_{b\uparrow}$.
$N_i = N_{i\uparrow} +N_{i\downarrow}$ is a constant. The last term
in ${\cal H}$ describes the spin-exchange scattering, which changes
the occupation of each spin state of each species and causes
interspecies  entanglement. Define $\mathbf{S}_a =
\sum_{\sigma,\sigma'} a_{\sigma}^{\dagger}
\mathbf{s}_{\sigma\sigma'}a_{\sigma'}$ and
$\mathbf{S}_b=\sum_{\sigma,\sigma'} b_{\sigma}^{\dagger}
\mathbf{s}_{\sigma\sigma'}b_{\sigma'}$, where $\mathbf{s}$ is the
spin-$\frac{1}{2}$ operator. The common eigenstate of $S_a^2$ and
$S_{az}$  is $|S_a,m_a\rangle =
|\frac{N_a}{2}+m_a\rangle_{a\uparrow}
|\frac{N_a}{2}-m_a\rangle_{a\downarrow}$. Thus we transform the
Hamiltonian into that of two coupled big spins $S_a= N_a/2$ and
$S_b=N_b/2$,
$$\begin{array}{ccl} \frac{{\cal H}}{J_z} & = &
\frac{K_e}{J_z}(S_{ax}S_{bx}+S_{ay}S_{by})+ S_{az} S_{bz}
\\&&
+ B_a S_{az} + B_b S_{bz} +C_a S_{az}^2 +C_b
S_{bz}^2+\frac{E_0}{J_z},
\end{array}$$
where
$J_z=K_{\uparrow\uparrow}^{(ab)}+K_{\downarrow\downarrow}^{(ab)}
-K_{\uparrow\downarrow}^{(ab)}-K_{\downarrow\uparrow}^{(ab)}$,
$B_aJ_z=f_{a\uparrow}-f_{a\downarrow}+N_a(K_{\uparrow\uparrow}^{(aa)}
-K_{\downarrow\downarrow}^{(aa)})/2+N_b(K_{\uparrow\uparrow}^{(ab)}
+K_{\uparrow\downarrow}^{(ab)} -K_{\downarrow\uparrow}^{(ab)}
-K_{\downarrow\downarrow}^{(ab)})/2$,
$B_bJ_z=f_{b\uparrow}-f_{b\downarrow}+N_b(K_{\uparrow\uparrow}^{(bb)}
-K_{\downarrow\downarrow}^{(bb)})/2+N_a(K_{\uparrow\uparrow}^{(ab)}
+K_{\downarrow\uparrow}^{(ab)}
-K_{\uparrow\downarrow}^{(ab)}-K_{\downarrow\downarrow}^{(ab)})/2$,
$C_aJ_z = (K_{\uparrow\uparrow}^{(aa)}
+K_{\downarrow\downarrow}^{(aa)}-K_{\uparrow\downarrow}^{(aa)}
-K_{\downarrow\uparrow}^{(aa)})/2$, $C_bJ_z =
(K_{\uparrow\uparrow}^{(bb)}
+K_{\downarrow\downarrow}^{(bb)}-K_{\uparrow\downarrow}^{(bb)}
-K_{\downarrow\uparrow}^{(bb)})/2$. $E_0=\sum_{i}(\sum_{\sigma}
f_{i\sigma})N_i+(1/8)\sum_{i}(\sum_{\sigma\sigma'}K_{\sigma\sigma'}^{(i)})
N_i^2+(1/4)(\sum_{\sigma\sigma'}K_{\sigma\sigma'}^{(ab)})N_aN_b$ is
a constant and is neglected henceforth.

${\cal H}$ conserves the total $z$-component spin
$S_z=(N_{a\uparrow}-N_{a\downarrow}+N_{b\uparrow}-N_{b\downarrow})/2$.
The total spin is  $S=S_a-S_b,\cdots,S_a+S_b$, letting $N_a \geq
N_b$. For a given $S_z$, any eigenstate of ${\cal H}$ can be written
as
$|\Psi\rangle = \sum_{m} g(m)|S_a,m\rangle|S_b,S_z-m\rangle$,
with  the summation range $max(-S_a,S_z-S_b) \leq m \leq
min(S_a,S_z+S_b)$.

At the isotropic point, i.e. if $K_e=J_z$ while $B_a=B_b=C_a=C_b=0$,
the Hamiltonian is simplified as
${\cal H}= J_z \mathbf{S}_a\cdot\nobreak  \mathbf{S}_b$,
whose eigenstates  are nothing but $|S,S_z\rangle$, for which $g(m)$
is the Clebsch-Gordon coefficient $\langle S_a,m,
S_b,S_z-m|S,S_z\rangle$.  The eigenvalues  are degenerate for
different values of $S_z$. There are $N_a-N_b+1$ degenerate ground
states $|G_{S_z}\rangle=|S_a-S_b,S_z\rangle$, which are found to be
exactly
\begin{equation}
|G_{S_z}\rangle = A(a_{\uparrow}^{\dagger})^{n_{\uparrow}}
(a_{\downarrow}^{\dagger})^{n_{\downarrow}}
(a_{\uparrow}^{\dagger}b_{\downarrow}^{\dagger}-
a_{\downarrow}^{\dagger}b_{\uparrow}^{\dagger})^{N_b}|0\rangle,
\label{g}
\end{equation}
where $n_{\uparrow}= N_a/2-N_b/2+S_z$, $n_{\downarrow}=
N_a/2-N_b/2-S_z$,
$A=[(N_a-N_b+1)!/(N_a+1)!N_b!(N_a/2-N_b/2+S_z)!(N_a/2-N_b/2-S_z)!]^{1/2}$
is the normalization factor. Particularly, for  $N_a=N_b=N$, the
ground state is uniquely
$|G_0\rangle
=(\sqrt{N+1}N!)^{-1}(a_{\uparrow}^{\dagger}b_{\downarrow}^{\dagger}-
a_{\downarrow}^{\dagger}b_{\uparrow}^{\dagger})^N|0\rangle$.

The non-mean-field nature of these ground states can be well
characterized by the  entanglement between the two species. Rewrite
$|G_0\rangle$ as $(\sqrt{N+1})^{-1}\sum_{m=0}^{N} (-1)^{N-m}
|m\rangle_{a\uparrow}|N-\nobreak m\rangle_{a\downarrow}|N-\nobreak
m\rangle_{b\uparrow} |m\rangle_{b\downarrow}$. The entanglement
between the occupation of each one of the four single particle
states and the rest of the system can be  quantified as the von
Neumann entropy of its reduced density matrix~\cite{shi}, which is
$\log_{N+1} (N+1)=1$, where we set the logarithmic base to be $N+1$
because for each single particle basis state, the occupation number
is $(N+1)$-valued. In a sense $|G_0\rangle$ is maximally entangled,
due to equal superposition of the $N+1$  quart-orthogonal  states.
As each species comprises two spin states, it can be seen that the
entropy of the entanglement between the total $a$ species and the
total $b$ species is given by the same entropy. For the general
eigenstate $|\Psi\rangle$  above,  one can also write it in terms of
occupation numbers in four single particle states, and obtain the
entanglement entropy as $-\sum_m g^2(m) \log_{N_b+1} g^2 (m)$, which
must be less than $1$ if $|\Psi\rangle \neq |G_0\rangle$, as
$g(m)^2$'s are not all equal for different $m$'s.

$|G_{S_z}\rangle$ is a condensation of interspecies pairs in the
same two-particle entangled state $\phi(\mathbf{r}_a,\mathbf{r}_b)
\equiv \frac{1}{\sqrt{2}}
[\phi_{a\uparrow}(\mathbf{r}_a)|\uparrow\nobreak\rangle_a
\phi_{b\downarrow}(\mathbf{r}_b)|\downarrow\nobreak\rangle_b
-\phi_{a\downarrow}(\mathbf{r}_a)|\downarrow\nobreak\rangle_a
\phi_{b\uparrow}(\mathbf{r}_b)|\uparrow\rangle_b]$, plus the extra
$a$-atoms. This can be seen by noting
$(a_{\uparrow}^{\dagger}b_{\downarrow}^{\dagger}-
a_{\downarrow}^{\dagger}b_{\uparrow}^{\dagger})^{N_b}= [\sqrt{2}\int
d^3r_a d^3r_b \psi_a^{\dagger}(\mathbf{r}_a)
\psi_b^{\dagger}(\mathbf{r}_b)
\phi(\mathbf{r}_a,\mathbf{r}_b)]^{N_b}$,
where  $\psi_a(\mathbf{r})=\sum_{\sigma}
a_{\sigma}\phi_{a\sigma}(\mathbf{r}_a)|\sigma\rangle_a$,
$\psi_b(\mathbf{r})=\sum_{\sigma}
b_{\sigma}\phi_{b\sigma}(\mathbf{r}_b)|\sigma\rangle_b$. Thus the
BEC order parameter, given by $\phi(\mathbf{r}_a,\mathbf{r}_b)$,  is
quantum-entangled~\cite{shi0}. Like Cooper pairing, the formation of
the  entangled two-particle state lowers the energy, as can be seen
by considering  one $a$-atom and one $b$-atom, with Hamiltonian
$h(\mathbf{r}_a)+h(\mathbf{r}_b) + U_1(\mathbf{r}_a-\mathbf{r}_b)
+U_2(\mathbf{r}_a-\mathbf{r}_b)
(|\uparrow\downarrow\nobreak\rangle\langle\uparrow\nobreak\downarrow\nobreak|
+|\downarrow\uparrow\nobreak\rangle\langle\downarrow\uparrow|)
$,  where $U_2 >0$.  Compared with the separable state
$\phi_{a}(\mathbf{r}_a)\phi_{b}(\mathbf{r}_b)
|\sigma\rangle|\sigma'\rangle$, the entangled two-particle state
$\phi_{a}(\mathbf{r}_a)\phi_{b}(\mathbf{r}_b)(|\uparrow\rangle
|\downarrow\nobreak\rangle-
|\downarrow\rangle|\uparrow\rangle)/\sqrt{2}$ lowers the energy by
$\int
U^{(ab)}_2(\mathbf{r}_a-\mathbf{r}_b)|\phi_{a}(\mathbf{r}_a)\phi_{b}(\mathbf{r}_b)|^2$.
For such energetic reason, our BEC occurs globally in the
interspecies entangled pair state $\phi(\mathbf{r}_a,\mathbf{r}_b)$,
rather than separately in the two species.

This entanglement can be experimentally probed by several means.
First, because of entanglement,  the particle number of each spin
state of each species is subject to strong fluctuation. For
$|G_0\rangle$, $\langle  N_{i\sigma} \rangle = N/2$, $\langle
N^2_{a\sigma} \rangle - \langle N_{i\sigma}\rangle^2 = N(N+2)/12$,
and thus $\sqrt{\langle N^2_{a\sigma}\rangle -\langle
N_{a\sigma}\rangle^2}/\langle N_{a\sigma}\rangle \approx
1/\sqrt{3}$. Note that the quantum mechanical average can only be
obtained through several runs of the measurement. Nevertheless, as
the density is a self-averaging quantity~\cite{altman}, one can
study, in a single image, the fluctuation of the density
$\rho_{i\sigma}(\mathbf{r}_i)=N_{i\sigma}|\phi_{i\sigma}(\mathbf{r}_i)|^2$.
Interestingly, $\sqrt{\langle \rho_{i\sigma}(\mathbf{r}_i)^2\rangle
-\langle \rho_{i\sigma}(\mathbf{r}_i)\rangle^2}/\langle
\rho_{i\sigma}(\mathbf{r}_i)\rangle = \sqrt{\langle
N^2_{i\sigma}\rangle -\langle N_{i\sigma}\rangle^2}/\langle
N_{i\sigma}\rangle$.

Secondly, this entanglement is also indicated by  nonvanishing of
the connected correlations between particle numbers of the two
species, $C_{\sigma,\sigma'} \equiv \langle
N_{a\sigma}N_{b\sigma'}\rangle - \langle N_{a\sigma}\rangle\langle
N_{b\sigma'}\rangle$. For $|G_0\rangle$, $C_{\sigma,\sigma}
=-N(N+2)/12$, $C_{\sigma,\bar{\sigma}} = N(N+2)/12$, where
$\bar{\sigma}\neq \sigma$. Again, these connected correlations can
be measured through the corresponding quantities of the densities,
$g(\mathbf{r}_a, \sigma; \mathbf{r}_b,\sigma') \equiv \langle
\rho_{a\sigma}(\mathbf{r}_a)\rho_{b\sigma'}(\mathbf{r}_b)\rangle -
\langle \rho_{a\sigma}(\mathbf{r}_a)\rangle\langle
\rho_{b\sigma'}(\mathbf{r}_b)\rangle =C_{\sigma,\sigma'}
|\phi_{a\sigma}(\mathbf{r}_a)\phi_{b\sigma}(\mathbf{r}_b)|^2$, and
$g(\mathbf{r}_a, \sigma; \mathbf{r}_b,\sigma')/\langle
\rho_{a\sigma}(\mathbf{r}_a)\rangle\langle
\rho_{b\sigma'}(\mathbf{r}_b)\rangle = C_{\sigma,\sigma'}/\langle
N_{a\sigma}\rangle\langle N_{b\sigma'}\rangle$.

These density fluctuations and correlations are those in the
original condensate, so it is not necessary to switch off the trap
to let the condensate freely expand. Nevertheless, one can do so in
order to obtain a larger image, which equally contains information
about the entanglement. This is because the free propagator does not
affect the entanglement. Moreover, the density fluctuations and
correlations in the expanded gas are proportional to corresponding
quantities in momentum space~\cite{altman}, which are related to
those of the particle numbers in a way similar to that in the
coordinate space, as described above.

Thirdly, this entanglement can also be probed by detecting atoms
which leave the trap. Consider two-atom measurements in which the
joint probability is obtained. In measuring the spin state of an
$a$-atom, the probability of obtaining $\sigma$ is $P_{i\sigma} =
\langle a_{\sigma}^{\dagger}a_{\sigma}\rangle/\sum_{\sigma'}\langle
a_{\sigma'}^{\dagger}a_{\sigma'}\rangle $~\cite{ashhab}. For a joint
measurement of the spins of an $a$-atom and a $b$-atom, the
probability of obtaining $\sigma$ for  the $a$-atom while  $\sigma'$
for the $b$-atom is $P_{\sigma,\sigma'} = \langle
b_{\sigma'}^{\dagger}a_{\sigma}^{\dagger}a_{\sigma}b_{\sigma'}\rangle/
\sum_{\sigma_a,\sigma_b}\langle
b_{\sigma_b}^{\dagger}a_{\sigma_a}^{\dagger}a_{\sigma_a}b_{\sigma_b}
\rangle.$ A non-entangled, or mean-field type,  state  is in the
form of $(\sqrt{N_1!N_2!N_3!N_4!})^{-1}
{a_{\mathbf{\hat{n}}}^{\dagger}}^{N_1}{a_{\mathbf{-\hat{n}}}^{\dagger}}^{N_2}
{b_{\mathbf{\hat{m}}}^{\dagger}}^{N_3}{b_{\mathbf{-\hat{m}}}^{\dagger}}^{N_4}|0\rangle,$
where $\mathbf{\hat{n}}$ and $\mathbf{\hat{m}}$ are two arbitrary
``directions''.  Hence for arbitrary chosen $\sigma_a$ and
$\sigma_b$, $P_{\sigma_a,\sigma_b} = P_{\sigma_a}P_{\sigma_b}$. In
contrast, for an entangled BEC considered here,
$P_{\sigma_a,\sigma_b}\neq P_{\sigma_a}P_{\sigma_b}.$ Indeed, for
$|G_0\rangle$, $P_{\sigma_a} = P_{\sigma_b} = 1/2$, but
$P_{\uparrow\downarrow}= P_{\downarrow\uparrow} = (2N+1)/6N$ while
$P_{\uparrow\uparrow}= P_{\downarrow\downarrow} = (N-1)/6N$.

Finally, this entanglement has feedback effects on the single
particle orbits. We can derive the equations governing the orbital
wavefunctions, as the counterparts of the Gross-Pitaevskii equation,
using the method of Ashhab and Leggett~\cite{lobo2}. The
minimization of the expectation value of ${\cal H}$ in the ground
state $|G_0\rangle$ gives rise to
$ \{-\frac{\hbar^2}{2m_a}\nabla^2+U_{a\sigma}(\mathbf{r})
+[N(N-1)/3]g^{(aa)}_{\sigma\sigma}
|\phi_{a\sigma}(\mathbf{r})|^2+[N(N-1)/6]g^{(aa)}_{\uparrow\downarrow}
|\phi_{a\bar{\sigma}}(\mathbf{r})|^2+[N(N-1)/6]g^{(ab)}_{\sigma\sigma}
|\phi_{b\sigma}(\mathbf{r})|^2
+[N(2N+1)/6]g^{(ab)}_{\sigma\bar{\sigma}}
|\phi_{b\bar{\sigma}}(\mathbf{r})|^2 \} \phi_{a\sigma}(\mathbf{r})
-[N(N+2)/12]g_e \phi^*_{b\bar{\sigma}}(\mathbf{r})
\phi_{b\sigma}(\mathbf{r})\phi_{a\bar{\sigma}}(\mathbf{r}) =
\mu_{a\sigma}\phi_{a\sigma}(\mathbf{r})$
and the likes for the other three wavefunctions,  where
$g^{(ij)}_{\sigma\sigma'}$ and $g_{e}$ are the coefficients in front
of the integrals in $K^{(ij)}_{\sigma\sigma'}$ and
$K_{e}$~\cite{scatter}. The  term proportional to $g_{e}$ is an
interference term, which persists even in the Thomas-Fermi limit. It
is absent in previous models and may lead to novel physical
properties.

We now study how this entanglement survives the coupling anisotropy
$J_z \neq K_e$ and the nonvanishing of $B_a$, $B_b$, $C_a$ and
$C_b$. We use the Lanczos method to numerically determine the ground
states for different parameter values. First we study how the
entanglement varies with $K_e/J_z$, setting $B_a=B_b=C_a=C_b=0$. As
shown in Figs.~1 and 2, the entanglement is maximized when
$K_e=J_z$, and is maximized at $S_z=0$. This maximum is $1$ if
$S_a=S_b$, and is less than $1$ otherwise. Roughly speaking, the
larger the spins, the sharper the deviation of the entanglement from
the maximum, but the details depend on all of $S_a$, $S_b$ and
$S_z$. The decrease for $K_e/J_z
>1 $ is much slower than the increase for $K_e/J_z <1$.

\begin{figure}
\rotatebox{-90}{\resizebox{4cm}{7cm}{
\includegraphics{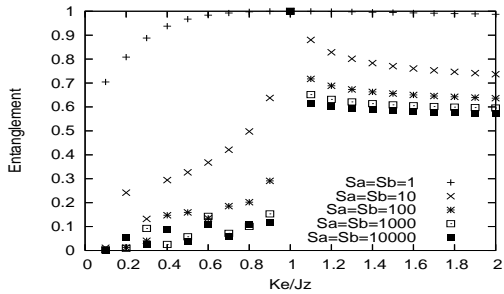}}} \caption{\label{pic1}The entanglement as a function
of $K_e/J_z$, with $B_a=B_b=C_a=C_b=0$, $S_z=0$. The entanglement
measure used both here and in the next three figures is the von
Neumann entropy of the reduced density matrix of either of the two
species, calculated by considering the occupation
entanglement~\cite{shi}.}
\end{figure}

\begin{figure}
\rotatebox{-90}{\resizebox{4cm}{7cm}{
\includegraphics{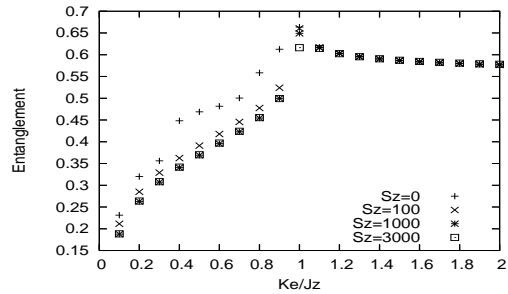}}}
\caption{\label{pic2}The entanglement as a function of $K_e/J_z$,
with $B_a=B_b=C_a=C_b=0$, $S_a=12000$, $S_b=10000$. }
\end{figure}

Under  typical values $S_a=12000$, $S_b=10000$, $S_z=1000$,
$K_e/J_z=1.2$, we study how the entanglement varies with $C_a$ and
$C_b$, while $B_a=B_b=0$. Because $J_z$, $C_aJ_z$ and $C_bJ_z$ are
about of the same order of magnitude of $K^{(ij)}_{\sigma
\sigma'}$'s,  we consider the range $-2 \leq C_a \leq 2$ and $-2
\leq C_b \leq 2$. As shown in Fig.~3, the change of entanglement
from the case of $C_a=C_b=0$, in which the entanglement is $0.6$, is
quite limited.

\begin{figure}
\rotatebox{-90}{\resizebox{4cm}{7cm}{
\includegraphics{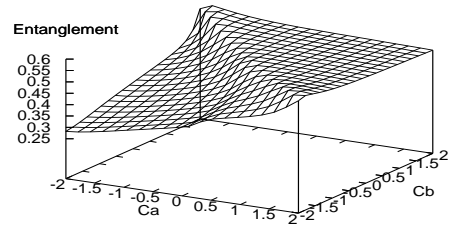}}}
\caption{\label{pic3} The entanglement as a function of $C_a$ and
$C_b$, with $S_a=12000$, $S_b=10000$, $S_z=1000$, $K_e/J_z=1.2$,
$B_a=B_b=0$. }
\end{figure}

Then we typically choose $C_a=0.2$ and $C_b=0.4$, and study the
variation of the entanglement with $B_a$ and $B_b$. As $B_a$ and
$B_b$ can be  as large as the order of magnitude of $N_a$ and $N_b$
or even larger, we consider the range $-10000 \leq B_a \leq 10000$
and $-10000 \leq B_b \leq  10000$. Fig.~4 indicates that in such a
parameter range, the variation of entanglement, with the maximum
$0.57$, is still quite limited.

\begin{figure}
\rotatebox{-90}{\resizebox{4cm}{7cm}{
\includegraphics{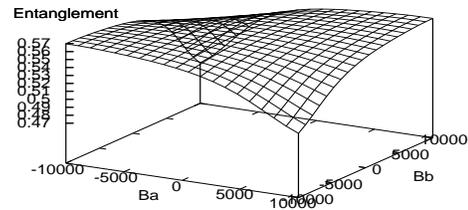}}}
\caption{\label{pic4} The entanglement as a function of $B_a$ and
$B_b$, with $S_a=12000$, $S_b=10000$, $S_z=1000$, $K_e/J_z=1.2$,
$C_a=0.2$, $C_b=0.4$. }
\end{figure}

These results show that  BEC with an inter-species entangled order
parameter, as a highly non-mean-field ground state, can occur in a
quite large parameter regime. The reason why the entanglement is not
easily destroyed by anisotropy and $C_a$, $C_b$, $B_a$ and $B_b$ is
that the entanglement is caused by the exchange interaction. This
leads to more relaxed condition for  the persistence of entanglement
than for the energetic perturbation. Moreover, around $C_a \approx
-C_b$, there is a canceling effect in the additional term in the
Hamiltonian, $C_aS_{az}^2+C_bS_{bz}^2 =
(C_a+C_b)S_{az}^2-2C_bS_zS_{bz}+C_bS_{z}^2$.  Similarly,  around
$B_a \approx B_b$, there is a canceling effect in
$B_aS_{az}+B_bS_{bz} = (B_a-B_b)S_{az}+B_{bz}S_z$. The canceling
effect further reduces the change of entanglement, as exhibited in
the Figs.~3 and 4.

It is important that the non-mean-field ground state, with
inter-species entanglement, can be accessed in experiments. A
crucial issue in experimentally accessing the non-mean-field ground
state is the energy gap. At the isotropic point of our Hamiltonian,
in analogy with BEC of spin-$1$ bosons,  the gap is of the order of
$\sim 1/V$, thus the ground state is very difficult to be reached in
practice~\cite{zhou}. This is spontaneous  symmetry breaking.
However, it can be estimated that when $B_a-B_b$ is of the order of
$\pm N$ or $C_a+C_b$ is of the order of $-1$, the energy gap is of
the order of $N/V$, which is finite in the thermodynamic limit. On
the other hand,  we have seen above that in this parameter regime,
as far as $K_e$ is larger than or not much smaller than $J_z$, the
entanglement is significantly nonvanishing. Even if $B_a$, $B_b$,
$C_a$ and $C_b$ are not so large, the finite energy gap can still be
opened up when $J_z \gg K_e $. But then the entanglement tends to
vanish.  One may resort to adiabatic switching, which might be
realized by using Feshbach resonance. One starts with the gapped
case $J_z \gg K_e$. By adiabatically tuning $K_e$ towards $K_e
\approx J_z$, the ground state evolves into a highly entangled one,
though with vanishing gap.

The hyperfine-Zeeman energy levels depend only on $F$ and $m_F$, not
on the species~\cite{leggett}. Therefore we can simply  represent
the same pseudo spin state of the two species by the same actual
hyperfine state. To exactly realize our model, only forward and
exchange scattering channels are needed, thus  the scattering into
other hyperfine states needs to be suppressed. A choice is
$|\uparrow\rangle = |F=2,m_F=2\rangle$, $|\downarrow\rangle =
|F=2,m_F=1\rangle$. One may also use $|\uparrow\rangle =
|F=2,m_F=2\rangle$, $|\downarrow\rangle = |F=1,m_F=1\rangle$,  or
use  $|\uparrow\rangle = |F=2,m_F=-1\rangle$, $|\downarrow\rangle =
|F=1,m_F=-1\rangle$, which are the spin states in the  well-known
two-component $^{87}Rb$ BECs~\cite{myatt}. In these previous
experiments, because the particle number in each spin state is
conserved, the BECs  are equivalent to a mixture of two species,
each in a fixed spin state, which was implemented in $^{41}K$ and
$^{87}Rb$, both in $|F=2,m_F=2\rangle$~\cite{modugno}. These
experimental experiences can exploited to realize our model. For the
latter two choices of spin states,  as the energy difference between
the two spin states with different $F$'s is very large, $B_a$ and
$B_b$ can be of the order of $N$ or larger. Likely this can produce
the gap. Lack of data on interspecies scattering prohibits more
detailed realistic discussions here.

To summarize, we have proposed a practically accessible
non-mean-field BEC, which is a condensation of interspecies
spin-entangled pairs and  can be characterized by a suitably defined
entanglement entropy, which is maximized in the ground state at the
isotropic limit, but remains significant in a wide parameter regime.
We have also derived the dynamical equations of the single particle
orbital wavefunctions, where there are novel interference terms.
This non-mean-field ground state can even be gapped. We have
discussed the experimental feasibility and put forward several means
of experimental detection.

Y.S. is very grateful to Prof. Tony Leggett for precious advice, and
to Fei Ye for generous help in numerics. We also thank Fei Zhou, Li
You, Michael Stone, Biao Wu and Tao Li for useful discussions.


\begin{thebibliography}{99}
\bibitem{scatter}
For the scattering in which an $i$-atom flips from $\sigma_4$ to
$\sigma_1$ while an $j$-atom flips from $\sigma_3$ to $\sigma_2$,
with scattering length
$\xi^{(ij)}_{\sigma_1\sigma_2\sigma_3\sigma_4}$ and effective mass
$\mu_{ij}=m_im_j/(m_i+m_j)$, define
$K^{(ij)}_{\sigma_1\sigma_2\sigma_3\sigma_4}\equiv
(2\pi\hbar^2\xi^{(ij)}_{\sigma_1\sigma_2\sigma_3\sigma_4}/\mu_{ij})\int
\phi_{i\sigma_1}^*(\mathbf{r})\phi_{j\sigma_2}^*(\mathbf{r})
\phi_{j\sigma_3}(\mathbf{r})\phi_{i\sigma_4}(\mathbf{r}) d^3r$.
$K^{(ii)}_{\sigma\sigma}\equiv K^{(ii)}_{\sigma\sigma\sigma\sigma}$,
$K^{(ii)}_{\sigma\bar{\sigma}}\equiv
2K^{(ii)}_{\sigma\bar{\sigma}\bar{\sigma}\sigma} =
2K^{(ii)}_{\sigma\bar{\sigma}\sigma\bar{\sigma}}$ for $\sigma\neq
\bar{\sigma}$, $K^{(ab)}_{\sigma\sigma'}\equiv
K^{(ab)}_{\sigma\sigma'\sigma'\sigma}$, $K_{e}\equiv
2K^{(ab)}_{\uparrow\downarrow\uparrow\downarrow}
=2K^{(ab)}_{\downarrow\uparrow\downarrow\uparrow}$.
\end{thebibliography}
\end{document}